\documentclass[a4paper,fleqn,usenatbib]{mnras}

\usepackage[T1]{fontenc}
\usepackage{ae,aecompl}

\usepackage{graphicx}	
\usepackage{amsmath}	
\usepackage{amssymb}	

 \usepackage[flushleft]{threeparttable}

\usepackage{color}

\newcommand\oH{\operatorname{\textit{o}-H}}
\newcommand\pH{\operatorname{\textit{p}-H}}

\title[Nitrogen fractionation revisited]{Revised Models of Interstellar Nitrogen Isotopic Fractionation}

\author[E. S. Wirstr\"{o}m \& S.B. Charnley]{
E. S. Wirstr\"{o}m,$^{1}$\thanks{E-mail: eva.wirstrom@chalmers.se}
and S. B. Charnley$^{2}$
\\
$^{1}$Department of Space, Earth and Environment, Chalmers University of Technology, Onsala Space Observatory, SE-439 92 Onsala, Sweden\\
$^{2}$Astrochemistry Laboratory, Code 691, NASA Goddard Space Flight Center, 8800 Greenbelt Road, Greenbelt, MD 20771, USA
}

\pubyear{2017}

\begin{document}
\label{firstpage}
\pagerange{\pageref{firstpage}--\pageref{lastpage}}
\maketitle

\begin{abstract}
Nitrogen-bearing molecules in cold molecular clouds exhibit a range of isotopic fractionation ratios and these molecules may be the precursors of $^{15}$N enrichments found in comets and meteorites. Chemical model calculations indicate that atom-molecular ion and ion-molecule reactions could account for most of the fractionation patterns observed. However, recent quantum-chemical computations demonstrate that several of the key processes are unlikely to occur in dense clouds. Related model calculations of dense cloud chemistry show that the revised $^{15}$N enrichments fail to match observed values. 
 
We have investigated the effects of these reaction rate modifications on the chemical model of Wirstr\"{o}m et al. (2012) for which there are significant physical and chemical differences with respect to other models. We have included  $^{15}$N fractionation of CN in neutral-neutral reactions and also updated rate coefficients for key reactions in the nitrogen chemistry.

We find that the revised fractionation rates have the effect of suppressing $^{15}$N enrichment in ammonia at all times, while the depletion is even more pronounced, reaching  $^{14}$N/$^{15}$N ratios of $>2000$. Taking the updated nitrogen chemistry into account, no significant enrichment occurs in HCN or HNC, contrary to observational evidence in dark clouds and comets, although the $^{14}$N/$^{15}$N ratio can still be below 100 in CN itself. However, such low CN abundances are predicted that the updated model falls short of explaining the bulk $^{15}$N enhancements observed in primitive materials. It is clear that alternative fractionating reactions are necessary to reproduce observations, so further laboratory and theoretical studies are urgently needed.
\end{abstract}

\begin{keywords}
astrochemistry -- molecular processes  -- ISM: molecules  -- ISM: clouds  --  meteorites, meteors, meteoroids  -- comets: general
\end{keywords}



\section{Introduction}
 
Accounting for the origin of the nitrogen isotopic enrichments  in primitive Solar System materials as measured in  meteorites, interplanetary dust particles and comets is a major problem in studies of the origin and evolution of the Solar System \citep{FuriMarty15,BockeleeMorvan15}.
Gaseous ion-molecule isotope exchange reactions at the low temperatures of the presolar molecular cloud have been recognized as a promising mechanism. \citet{TerzievaHerbst00} computed the  energetics of several viable processes  and these have formed the basis of several subsequent theoretical studies \citep{CharnleyRodgers02,RodgersCharnley08,RodgersCharnley08_ApJ,Hily-Blant13a,Wirstrom12,Heays14}. 
As summarized by \citet{Wirstrom16_FM}, when compared to observations of molecular clouds and Solar System objects, until recently these models have been successful in explaining some of  the measured $^{14}$N/$^{15}$N ratios. Assuming a nominal elemental $^{14}$N/$^{15}$N ratio of 400 for the ISM (cf. 440 in the Sun) the low HC$^{14}$N/HC$^{15}$N and H$^{14}$NC/H$^{15}$NC  ratios of cold molecular clouds  ($\approx$ 140--366)  and  low-mass protostars  ($\approx$ 160--460) could be reproduced. Based on the fractionation chemistry of \citet{TerzievaHerbst00} the reaction
 \begin{equation}
 {\rm ^{15}N ~+~ HC^{14}NH^+   } ~\leftrightharpoons~ {\rm ^{14}N ~+~ HC^{15}NH^+  } 
\label{eq:HCNH}  
\end{equation}
 underlies the  $^{15}$N enrichment of HCN and HNC. However, HCNH$^+ $ recombination with electrons produces CN radicals as well as HCN and HNC. The fact that CN appears to be generally less fractionated  \citep{Hily-Blant13b,Fontani15} is therefore puzzling. \citet{TerzievaHerbst00} also proposed that molecular nitrogen could become enriched in $^{15}$N through
  \begin{eqnarray}
{\rm  ^{15}N  ~+~ ^{14}N_2H^+ } ~\leftrightharpoons~ {\rm  ^{14}N  ~+~ ^{14}N^{15}NH^+} \\
 {\rm  ~~~~~~~~~~~~~~~~~~~~ } ~\leftrightharpoons~ {\rm  ^{14}N  ~+~ ^{15}N^{14}NH^+}  \label{eq:third}
\end{eqnarray}

  Model calculations indicate a similar $^{15}$N enrichment  of   ${\rm N_2H^+ } $ isotopologues as for  the nitriles \citep[e.g.][]{Wirstrom12}. However, recent studies of dark clouds and regions of massive star formation by \citet{Bizzocchi13} and \citet{Fontani15} indicate that, while the predicted range of enrichments are indeed evident, dramatic depletions  ($>$1000) are also found. 
On the other hand,  the $^{14}$NH$_3$/$^{15}$NH$_3$ ratios observed \citep{Gerin09, Lis10, Adande17} and calculated \citep{Wirstrom12, Hily-Blant13a} do not appear to show such a large range of values.  In the relatively  few sources where the isotopologues of {\it both} ammonia and ${\rm N_2H^+ } $ have been detected, the  $^{14}$N/$^{15}$N ratios do seem to roughly correlate \citep{Wirstrom16_FM,Adande17}, consistent with ammonia being formed from N$_2$.
When compared to the cometary  $^{14}$N/$^{15}$N ratios these models can explain the measured HCN and CN ratios (assuming that the later is derived from HCN) but fail to account for the large ammonia enrichments 
\citep[$\approx~ 90-190$,][]{BockeleeMorvan15,Shinnaka16}.  There are no data on cometary N$_2$ fractionation. 

The veracity of existing theoretical fractionation models for molecular $^{14}$N/$^{15}$N ratios,  and their comparison with the meteoritic record, comets and the interstellar medium,   have now been challenged in a recent study by \citet{Roueff15} (hereafter R15). Roueff et al. developed a chemical model for $^{15}$N and $^{13}$C fractionation in dense molecular clouds in which reaction rates for the nitrogen fractionation reactions of \citet{TerzievaHerbst00} were re-evaluated based on theoretical calculations and updated zero- point energy (ZPE) values from recent spectroscopic data. 
   Most importantly, the quantum chemical calculations presented by Roeuff et al.  indicate that barriers in the entrance channels will render reactions (\ref{eq:HCNH})--(\ref{eq:third}) ineffective,  preventing efficient fractionation of N$_2$ as well as the nitriles \citep{RodgersCharnley08}. 
   
The results of R15  will also have a detrimental effect on the fractionation calculations reported by  \citet{Wirstrom12} (hereafter W12). Although it could not account for  the depletions observed in the ${\rm N_2H^+ } $ isotopologues, the W12  model was successful in explaining several other aspects of observed interstellar $^{15}$N fractionation - enriched nitrile fractionation; modest ammonia fractionation -  and, through the evolution of the H$_2$ ortho to para spin ratio (OPR),  the correlations and anti-correlations between D and $^{15}$N found in primitive materials \citep{vanKooten17}. 

In this paper we report calculations that re-evaluate those of W12 in light of the revised R15 fractionation chemistry. This study is necessary because, apart from the rate coefficients adopted,  there exist significant differences in the physics and chemistry of the models of W12 and R15. 
 The  calculations of W12 considered densities similar to pre-stellar core conditions, about 10--100 times greater than those of R15 which were based on a standard dark cloud similar to TMC-1 ($n_{\rm H}$ = 0.2--2$\times10^5$ cm$^{-3}$). In contrast to W12, R15  neither included the depletion of gas-phase species by freeze-out onto dust grains, nor the time-dependence in the H$_2$ OPR -- both effects which have been shown to have substantial impact on the relative fractionation in nitriles and amines (here nitrogen hydrides and chemical derivatives thereof) over time (W12). 

     Furthermore, even when reactions (\ref{eq:HCNH})--(\ref{eq:third}) are not operating, there are alternative fractionation processes that could still become important in the physical conditions of the W12 model.  For example,  W12 identified 
  \begin{equation}
 {\rm ^{15}N^+  ~+~  ^{14}N_2    } ~\leftrightharpoons~ {\rm ^{14}N^+  ~+~ ^{14}N^{15}N   } 
\label{eq:Nplus}  
\end{equation}
 as playing a role in the fractionation of N$_2$ and  NH$_3$, whereas \citet{RodgersCharnley08} speculated that the neutral-neutral pathway 
  \begin{equation}
 {\rm ^{15}N  ~+~  C^{14}N  } ~\leftrightharpoons~ {\rm ^{14}N  ~+~  C^{15}N   } 
\label{eq:SDR1}  
\end{equation}
could be viable and efficient if there is a barrier to the reaction 
  \begin{equation}
 {\rm N  ~+~  CN  } ~\longrightarrow~ {\rm N_2  ~+~  C.   } 
\label{eq:SDR2}  
\end{equation}

However, the reaction appears to have no formation barrier, and the recommended reaction rate from the KIDA database \citep{Wakelam13} of $k({\rm T}) = 8.8\times10^{-11}\,({\rm T}/300)^{0.42}$~cm$^3$\,molecule$^{-1}$\,s$^{-1}$  agrees with laboratory  measurements down to 56~K \citep{Daranlot12}.   W12 assumed a 200~K barrier for reaction (\ref{eq:SDR2}) \citep[see][]{LeTeuff00} and  R15  found that isotopic exchange through (\ref{eq:SDR1}) was indeed plausible, a process not considered in W12.
We  have therefore updated the W12 model with respect to the revisions to nitrogen fractionation advocated by  R15, and specific reactions in the nitrogen chemistry from the KIDA database \citep{Wakelam13}. 

 The paper is structured as follows.
Revisions to the $^{14}$N fractionation chemistry and the updates to specific reactions in the nitrogen chemistry are are described in $\S 2$. The chemical model is summarised in  $\S 3$.  Results from the model calculations are given in  $\S 4$ where they are compared to previous work. Conclusions from this study are in  $\S 5$.

\section{Revised Chemical Network}
 
\subsection{$^{15}$N fractionation} 
 
\begin{table*}
	\centering
	\caption{Nitrogen isotope fractionation reactions} \label{ReacTab}
	\begin{threeparttable}
	\begin{tabular}{llccc} 
		\hline
		No. & Reaction & $\Delta E/k$ & $K(T)_{\rm W12}$ & $K(T)_{\rm upd}$ \\
		& & (K) & (10~K) & (10~K) \\
		\hline
		RF1a & ${\rm N_2H^+\,+\,N^{15}N}~\rightleftharpoons~{\rm N^{15}NH^+\,+\,N_2}$  & 10.3 & 1.46 & 1.40  \\
		RF1b & ${\rm N_2H^+\,+\,N^{15}N }~\rightleftharpoons~{\rm^{15}NNH^+\,+\,N_2}$ & \phantom{0}2.1 & 0.63 & 0.62 \\
		RF2 & ${\rm^{15}NNH^+\,+\,N^{15}N}~\rightleftharpoons~{\rm N^{15}NH^+\,+\,N^{15}N}$ & \phantom{0}8.1 & -\tnote{a} & 2.25 \\
		RF3 & ${\rm N_2H^+\,+\,^{15}N_2}~\rightleftharpoons~{\rm ^{15}N_2H^+\,+\,N_2}$ & 15.0\tnote{c} & 4.48 & 4.48 \\
		RF4 & ${\rm N^{15}NH^+\,+\,^{15}N_2}~\rightleftharpoons~{\rm ^{15}N_2H^+\,+\,N^{15}N}$  & 4.3\tnote{c} & 3.07 & 3.07 \\
		RF5 & ${\rm ^{15}NNH^+\,+\,^{15}N_2}~\rightleftharpoons~{\rm ^{15}N_2H^+\,+\,N^{15}N}$ &12.7\tnote{c} & 7.12 & 7.12 \\\\

		RF6a &${\rm N_2H^+\,+\,^{15}N}~\rightleftharpoons~{\rm N^{15}NH^+\,+\,N}$  & 38.5   & 37.0  & -\tnote{b} \\
		RF6b &${\rm N_2H^+\,+\,^{15}N}~\rightleftharpoons~{\rm ^{15}NNH^+\,+\,N}$  & 30.4   & 16.0  & -\tnote{b} \\
		RF7 & ${\rm HCNH^+\,+\,^{15}N}~\rightleftharpoons~{\rm HC^{15}NH^+\,+\,N}$ & 37.1 & 36.2 &  -\tnote{b} \\
		
		RF8 & ${\rm CNC^+\,+\,^{15}N}~\rightleftharpoons~{\rm C^{15}NC^+\,+\,N}$ & 38.1 & 38.1 & 45.2 \\
		RF9 & ${\rm CN\,+\,^{15}N}~\rightleftharpoons~{\rm C^{15}N\,+\,N}$ & 22.9 & -\tnote{a} & \phantom{0}9.9 \\\\
		
		RF10 & ${\rm ^{15}N^+\,+\,N_2}~\rightleftharpoons~{\rm N^+\,+\,N^{15}N}$ & 28.3 & 16.9 & 33.9 \\
		RF11 & ${\rm ^{15}N^+\,+\,NO}~\rightleftharpoons~{\rm N^+\,+\,^{15}NO}$ & 24.3 & 11.4  & -\tnote{b} \\\\
		RF12 & ${\rm ^{15}N\,+\,C_2N}~\rightleftharpoons~{\rm N\,+\,C_2\!^{15}N}$ & 26.7 & -\tnote{a}  & 14.4\tnote{d} \\
		\hline
	\end{tabular}
	\begin{tablenotes}
            \item[a] Not included in the W12 network.
            \item[b] A reaction barrier prevents this reaction from taking place at 10~K \citep{Roueff15}.
            \item[c] Not evaluated by \citet{Roueff15}, but from \citet{AdamsSmith81}.
            \item[d] Only included in the W12+R15+W13 network.
        \end{tablenotes}
        \end{threeparttable}
\end{table*}


\begin{table*}
	\centering
	\caption{Revised rates for nitrogen chemistry} \label{KIDATab}
	\begin{threeparttable}
	\begin{tabular}{llccc} 
		\hline
		No. & Reaction & $k_{\rm W12}{\rm (10~K)}$ & $k_{\rm upd}{\rm (10~K)}$ \\
		& & ($\rm cm^3s^{-1}$) & ($\rm cm^3s^{-1}$)  \\
		\hline
RN1a& ${\rm C\,+\,NH_2 ~\longrightarrow~ HNC\,+\,H}$ & 3.65$\times 10^{-12}$ & 3.25$\times 10^{-11}$\\
RN1b& ${\rm C\,+\,NH_2 ~\longrightarrow~ HCN\,+\,H}$ & 0.0\tnote{a} & 3.25$\times 10^{-11}$\\
RN2a& ${\rm CH\,+\,NH_3 ~\longrightarrow~ HCN\,+\,\oH_2\,+\,H}$ & 0.0\tnote{a} & 4.74$\times 10^{-12}$\\
RN2b& ${\rm CH\,+\,NH_3 ~\longrightarrow~ HCN\,+\,\pH_2\,+\,H}$ & 0.0\tnote{a} & 4.74$\times 10^{-12}$\\
RN3a& ${\rm N\,+\,CH_2 ~\longrightarrow~ HCN\,+\,H }$ & 3.65$\times 10^{-12}$ &  2.80$\times 10^{-11}$ \\
RN3b& ${\rm N\,+\,CH_2 ~\longrightarrow~ HNC\,+\,H}$ & 0.0\tnote{a} &  2.80$\times 10^{-11}$ \\
RN4 & ${\rm N\,+\,NH ~\longrightarrow~ N_2\,+\,H}$ & 4.98$\times 10^{-11}$ & 3.56$\times 10^{-11}$\\
RN5& ${\rm N\,+\,NH_2 ~\longrightarrow~ N_2\,+\,H\,+\,H}$ & 0.0\tnote{a} & 1.20$\times 10^{-10}$ \\
RN6& ${\rm N\,+\,C_2N ~\longrightarrow~ CN\,+\,CN }$ & 0.0\tnote{b} & 1.00$\times 10^{-10}$ \\
RN7 & ${\rm N\,+\,OH ~\longrightarrow~ NO\,+\,H }$ & 1.25$\times 10^{-10}$ &  2.74$\times 10^{-11}$\\
RN8 & ${\rm N\,+\,CN ~\longrightarrow~ N_2\,+\,C}$ & 6.18$\times 10^{-19}$\tnote{c} & 2.11$\times 10^{-11}$\\
RN9 & ${\rm N\,+\,NO ~\longrightarrow~ N_2\,+\,O}$ & 3.10$\times 10^{-11}$ & 1.07$\times 10^{-11}$\\
RN10a & ${\rm O\,+\,NH ~\longrightarrow~ NO\,+\,H}$ & 1.16$\times 10^{-10}$ &  6.60$\times 10^{-11}$\\
RN10b& ${\rm O\,+\,NH ~\longrightarrow~ OH\,+\,N}$ & 1.16$\times 10^{-11}$ & 0.0 \\
RN11 & ${\rm O\,+\,NH_2 ~\longrightarrow~ NH\,+\,OH}$ & 1.16$\times 10^{-11}$ & 9.84$\times 10^{-12}$\\
RN12 & ${\rm O\,+\,CN ~\longrightarrow~ CO\,+\,N}$ &  1.15$\times 10^{-13}$ & 5.00$\times 10^{-11}$\\
RN13a & ${\rm CN\,+\,NH_3 ~\longrightarrow~ HCN\,+\,NH_2}$ & 1.10$\times 10^{-09}$ &  5.04$\times 10^{-10}$ \\
RN13b & ${\rm CN\,+\,NH_3 ~\longrightarrow~ NH_2CN\,+\,H }$ & 1.00$\times 10^{-13}$ & 0.0 \\
RN14a & ${\rm NH_3^+\,+\,\oH_2 ~\longrightarrow~ NH_4^+\,+\,H}$ & 2.00$\times 10^{-12}$ & 1.19$\times 10^{-12}$ \\
RN14b& ${\rm NH_3^+\,+\,\pH_2 ~\longrightarrow~ NH_4^+\,+\,H}$ & 2.00$\times 10^{-12}$ & 1.19$\times 10^{-12}$ \\
RN15a & ${\rm N_2H^+\,+\,e^- ~\longrightarrow~ N_2\,+\,H}$ & 3.63$\times 10^{-6}$ & 4.30$\times 10^{-6}$\\
RN15b & ${\rm N_2H^+\,+\,e^- ~\longrightarrow~ NH\,+\,N}$ & 0.0\tnote{a} & 2.26$\times 10^{-7}$\\
RN16 & ${\rm NH^+\,+\,e^- ~\longrightarrow~ N\,+\,H}$ & 2.36$\times 10^{-7}$ & 2.11$\times 10^{-6}$ \\
RN17a & ${\rm N^+\,+\,\oH_2 ~\longrightarrow~ NH^+\,+\,H}$ & 8.74$\times 10^{-12}$	 & 7.97$\times 10^{-12}$ \\
RN17b & ${\rm N^+\,+\,\pH_2 ~\longrightarrow~ NH^+\,+\,H}$ & 4.02$\times 10^{-17}$	 &1.64$\times 10^{-19}$ \\
RN18a & ${\rm ^{15}N^+\,+\,\oH_2 ~\longrightarrow~ ^{15}NH^+\,+\,H}$ & 8.74$\times 10^{-12}$	& 1.32$\times 10^{-11}$ \\
RN18b & ${\rm ^{15}N^+\,+\,\pH_2 ~\longrightarrow~ ^{15}NH^+\,+\,H}$ & 4.02$\times 10^{-17}$	& 2.87$\times 10^{-19}$ \\
		 		\hline
	\end{tabular}
	\begin{tablenotes}
            \item[a] Not included in the W12 network.
            \item[b] The C$_2$N species was not included in the W12 network.
            \item[c] \citet{RodgersCharnley08} considered models with zero rate coefficient and W12 used a 200~K energy barrier. 
        \end{tablenotes}
        \end{threeparttable}
\end{table*}

Table~\ref{ReacTab} lists the major reactions potentially leading to $^{15}$N fractionation in cold, dark clouds, both ion-molecule and neutral-neutral reactions, with ZPE differences, $\Delta E$, as determined by R15. The resulting equilibrium coefficient, $K(T) = k_{\rm f} / k_{\rm r}$, the ratio between the forward and reverse reaction rate \citep[e.g. ][]{TerzievaHerbst00}, is listed for each reaction at 10~K, both for the new fractionation network and the one from W12.
Comparing the equilibrium coefficients in Table~\ref{ReacTab}, there are three major differences between the two networks: 
\begin{itemize}
 
\item (i) Atomic $^{15}$N is only contributing directly to the fractionation of molecules through reactions (RF8) and (RF9), while three of the most important fractionation reactions in previous models, (RF6a), (RF6b), and (RF7), have been found unlikely to proceed at 10~K. \textit{This is likely to suppress the fractionation in all nitriles as well as N$_2$H$^+$.}  
\item (ii) A fractionation in N$_2$ is introduced in the proton transfer reaction (RF2). \textit{This is likely to only shift the relative abundance of $^{15}$NNH$^+$ and N$^{15}$NH$^+$.} 
\item (iii) The atomic ion $^{15}$N$^+$ will no longer fractionate NO (RF11) and is more efficiently circulated back into molecular nitrogen through (RF10). \textit{Since $^{15}$N$^+$ is initiating all nitrogen fractionation in ammonia (as discussed in W12), this effect is expected to be suppressed.}
\end{itemize}

The revised reaction data of (RF1)--(RF11) were incorporated into the W12 chemical network with the rates determined by R15 (the 10~K upper limit given for the reaction rate of reaction (RF9) was adopted: 2.1$\times 10^{-11} \rm~cm^3\,s^{-1}$). 
The resulting reaction network, W12+R15, has 4437 reactions between 281 gas-phase species. 

\subsection{Nitrogen chemistry} 

Many of the reactions describing and dominating the nitrogen chemistry in a cold molecular cloud have not been studied in detail, but are assumed to proceed at approximate reaction rates, e.g. the classical Langevin collision rate for ion-molecule reactions. Through sensitivity analysis, \citet{Wakelam10b} identified 18 neutral-neutral and 2 ion-molecule nitrogen reactions to which abundance determinations are particularly sensitive, and \citet{Wakelam13} present a theoretical review of these reactions, their rates and branching ratios. Following recommendations by \citet{Wakelam13} for rates valid down to 10 K in temperature, 15 reaction channels are modified, and seven reactions or reaction channels are added to the nitrogen reaction network of W12.   
The relevant reactions are listed as (RN1--15) in Table~\ref{KIDATab}. 

The C$_2$N species was not included in the W12 network, but as its reaction with atomic nitrogen, (RN6), can be an important path to CN, it has been included in this updated network together with its protonated form, HC$_2$N$^+$, and their $^{15}$N isotopologues. The formation and destruction paths were adopted from the KIDA database \citep{Wakelam15}, and the fractionation reaction (RF12) (Table~\ref{ReacTab}) included.

In addition, the electron dissociative recombination of NH$^+$ was investigated experimentally and found to proceed at significantly higher rates at 10~K than previously assumed \citep{Novotny14}, so the rate has been updated (RN16). 
Corresponding reactions for $^{15}$N isotopologues are assumed to have identical reaction rates to those for the main isotopologue, and, when applicable, reactions are  assumed to produce ortho and para versions of H$_2$ with equal probability. The only exception is the reaction N$^+$ + \textit{o}/\textit{p}-H$_2$, for which \citet{Grozdanov16} recently calculated state- and isotope- specific rates (RN17,RN18) which are applied. These are expected to suppress the formation of amines somewhat and affect their isotopic ratios, since the reaction rate for forming $^{15}$NH$^+$ is found to be more than 60 per cent higher than for forming $^{14}$NH$^+$. While the Grozdanov rates do not reproduce experimental results for the \textit{p}-H$_2$ reactions as well as the rates calculated by \citet{Dislaire12} (used in W12), inclusion of the potential fractionation effects can be important and the effect on overall ammonia abundance should be very limited.

The resulting network is designated W12+R15+W13, and includes 4495 reactions between 285 gas-phase species.

\section{Chemical model}
 
Following W12, we consider the chemical evolution of the central regions of a static prestellar core with a gas density $n$(H$_2$)=10$^6$~cm$^{-3}$, a temperature of 10~K and a visual extinction of $A_{\rm V}$$>$10~mag. Cosmic ray ionization occurs, at a rate of $\zeta $=3$\times$10$^{-17}$~s$^{-1}$. The chemical model used is based on \citet{RodgersCharnley08} and W12, including the total expansion of $ortho$ and $para$ forms of H$_2$, H$_2^+$, and H$_3^+$. An initial H$_2$ OPR of 3 is assumed, but this decreases over time through the conversion reaction 
  \begin{equation}
 {\rm H^+  ~+~  {\it o}-H_2   ~\longrightarrow~ {\rm {\it p}-H_2  }  ~+~  H^+    } 
\label{eq:H+H2}  
\end{equation}
 which is exothermic by 170.5~K. Since the model includes the continuous production of H$_2$ at OPR=3 \citep{FukutaniSugimoto13} from atomic H sticking to grain surfaces with an efficiency of 0.6, an equilibrium gas-phase OPR is eventually reached. Apart from H$_2$ molecules leaving the grains upon formation, no grain surface chemistry or desorption is considered. However, all neutral gas-phase species, except H$_2$, He, N, and N$_2$, stick and freeze out onto grains upon collision. 

Helium, carbon, oxygen and nitrogen are included at elemental abundances  given by \citet{SavageSembach96}. We are interested in cores that are just about to form protostars and so we assume that all carbon is initially bound up in CO, with the remaining oxygen in atomic form, and the nitrogen is partly atomic with 50 per cent in molecular and atomic form respectively.
The elemental $^{14}$N/$^{15}$N ratio is assumed to be 440, with the same fraction of $^{15}$N in atomic form initially, $^{15}$N/$^{15}$N$^{14}$N=1. Note that this implies a nominal N$_2$/N$^{15}$N ratio of 220.

\section{Results} \label{ResSec}
 
\subsection{$^{15}$N fractionation revisions}
\begin{figure}
	\includegraphics[width=\columnwidth]{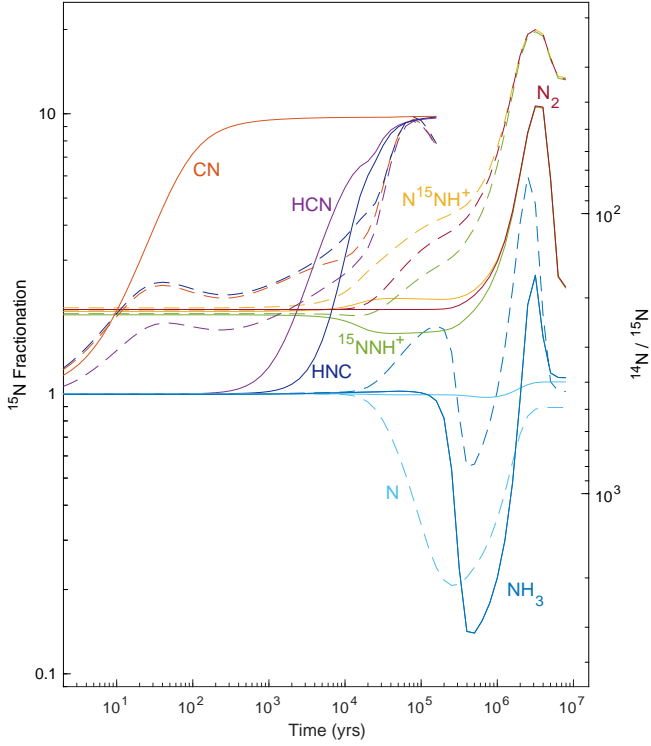}
    \caption{Nitrogen fractionation in the major nitrogen-bearing gas-phase species, relative to an elemental $^{15}$N/$^{14}$N ratio of 1/440, as predicted by the original W12 model (dashed) and the updated W12+R15 model (solid). The right-hand axis scale show the corresponding $^{14}$N/$^{15}$N ratio. A colour version of this figure is found in the electronic journal. }
    \label{FigRoueffFrac}
\end{figure}

\begin{figure}
	\includegraphics[width=\columnwidth]{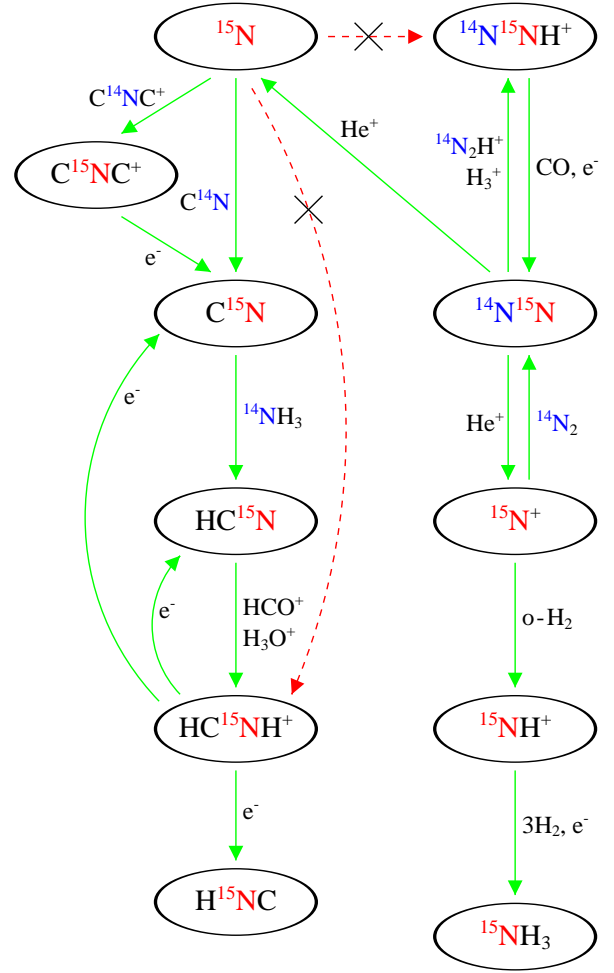}
    \caption{Revised chemical network illustrating the main reactions (green, solid arrows) responsible for $^{15}$N fractionation in nitriles and amines in the W12+R15 model. Red, dashed arrows mark fractionation reactions which are ineffective according to R15.}
    \label{FigFracNetw}
\end{figure}

Figure~\ref{FigRoueffFrac} shows the evolution of $^{15}$N fractionation in the major nitrogen species as calculated from model W12+R15 as compared to W12, and Fig.~\ref{FigFracNetw} illustrates the most important fractionating reactions. In what follows we cite reactions directly to Tables 1 and 2.
Contrary to predictions by R15, an enhancement of a factor of $\sim10$ is built up in the nitriles, as in W12. All available atomic $^{15}$N rapidly fractionates CN directly through reaction (RF9), and through (RF8) followed by dissociative electron recombination. Figure~\ref{FigFracNetw} shows how this $^{15}$N enhancement is successively spread to HCN, e.g. through the reaction 
  \begin{equation}
 {\rm CN  ~+~ NH_3  } ~\longrightarrow~ {\rm HCN  ~+~  NH_2   } 
\label{eq:CNammonia}  
\end{equation}
and even later to HNC through HCNH$^+$, formed by reactions of HCN with e.g. HCO$^+$ and ${\rm H_3O^+}$. 

The fractionation in ammonia follows closely the atomic ion ratio, ${\rm ^{15}N^+/^{14}N^+}$, since its formation is initiated by reactions (RN16,17) see e.g. W12 and \citet{Grozdanov16}. 
However, NH$_3$ is not enhanced in $^{15}$N as CO is depleted from the gas-phase after about 10$^4$ years, as found in W12, because reaction (RF10) channels ${\rm ^{15}N^+}$ into ${\rm ^{14}N^{15}N}$ faster than into ${\rm ^{15}NH^+}$ (see Fig.~\ref{FigFracNetw}). Consequently, as the \textit{o}-H$_2$ abundance drops after $2\times10^5$~yrs, the $^{15}$N depletion in ammonia becomes even more pronounced than in W12, and the later enhancement 
suppressed. 
The molecular reservoir of nitrogen remains unfractionated until late in core evolution, $\sim10^6$~yrs. Unlike earlier model predictions, including those of W12 and R15, ${\rm ^{15}NNH^+}$ is slightly reduced in $^{15}$N from about 10$^4$ to $5\times10^5$~yrs (${\rm^{14}N/^{15}N}$ of about 300 compared to the nominal ratio of 220, see Fig.~\ref{FigRoueffFrac}), but far from the very depleted ratios observed in L1544 \citep{Bizzocchi13}. The reason for this decrease is that in the absence of reactions (RF6), the formation and destruction of diazenylium (N$_2$H$^+$) are dominated by non-fractionating reactions from early times: production by H$_3^+$ + N$_2$ and destruction by CO back to N$_2$, so that the $^{15}$N fractionation simply follows the molecular ${\rm N_2/N^{15}N}$ ratio. As CO is depleted from the gas-phase, N$_2$ becomes successively more important in the destruction of ${\rm ^{15}NNH^+}$, effectively converting it to N$_2$H$^+$ (the reverse of reaction RF1b) and thus increasing the 
$\rm N_2H^+/^{15}NNH^+ $. Formation and destruction of ${\rm N^{15}NH^+}$ follows a similar pattern, but the net effect of the fractionating reactions (RF1) and (RF2) is instead a slight enhancement in $^{15}$N relative to N$_2$, see Fig.~\ref{FigRoueffFrac}.

\subsection{Nitrogen chemistry revisions}   
 
\begin{figure*}
	\includegraphics[width=\textwidth]{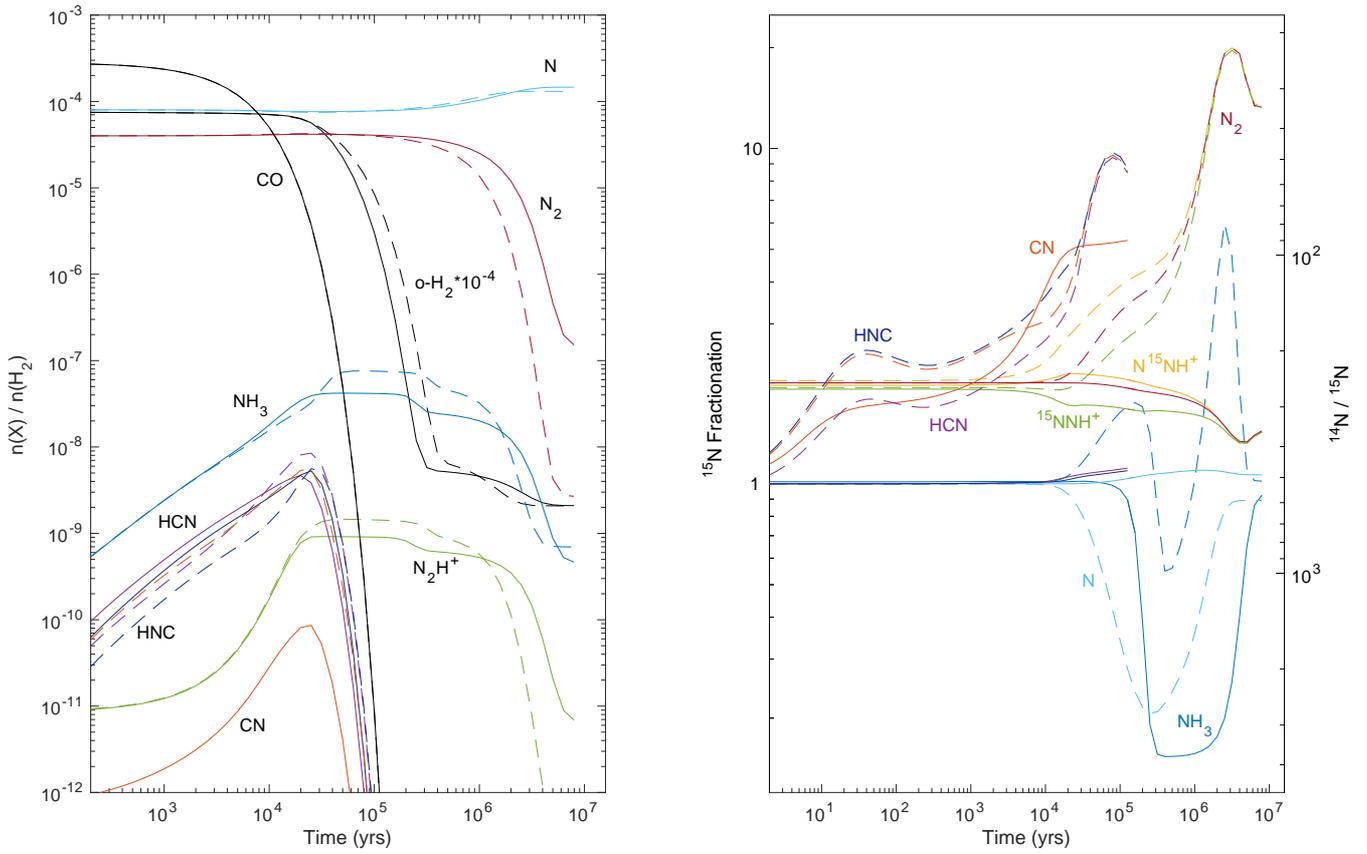}
    \caption{Left panel: Time evolution of the nitrogen chemistry in dense cores for nitrogen reaction rates based on \citet{Wakelam13} (solid lines), as compared to the W12 nitrogen reaction network (dashed lines). Right panel: Predicted evolution of the nitrogen fractionation in the major nitrogen-bearing gas-phase species for the combined W12+R15+W13 chemical network (solid lines) as compared to the W12 network (dashed lines). On the left-hand axis $^{15}$N enhancements are given relative to an elemental $^{15}$N/$^{14}$N ratio of 1/440, while the scale of the right-hand axis show the corresponding $^{14}$N/$^{15}$N ratio. A colour version of this figure is found in the electronic journal. }  
    \label{FigKIDARoueffAbuFrac}  
\end{figure*}

In Figure~\ref{FigKIDARoueffAbuFrac}, the abundance and $^{15}$N fractionation evolution of the major species using the updated W12+R15+W13 network are compared to results from the W12 model. From the left panel it is evident that almost two orders of magnitude less CN is maintained in the gas-phase of a dense core when applying the nitrogen reaction rates of Table~\ref{KIDATab}. We note that the inclusion of reaction (RN6) does not have a significant effect on the overall CN abundance under the present conditions. Since nitriles are mainly enhanced in $^{15}$N from reactions with CN in the W12+R15 fractionation network (see Fig.~\ref{FigFracNetw}), this low CN abundance also has the effect of suppressing $^{15}$N fractionation in HCN. Thus, the previously predicted enhancements of HC$^{15}$N  and H$^{15}$NC are almost completely eliminated, even though CN itself still reaches enhancements corresponding to CN/C$^{15}$N$\sim$80 (right panel of Fig.~\ref{FigKIDARoueffAbuFrac}). The difference causing this large discrepancy is of course that reaction (RN8) had effectively zero reaction rate at 10~K in the W12 model (see $\S 1$). Note that abundance and fractionation of HNC follow closely that of HCN, typically at slightly lower values.

Another effect of the higher efficiency of reaction (RN8) is that N$_2$ is maintained in the gas-phase later in cloud evolution, which also suppress the late increase in $^{15}$N fractionation in N$_2$, N$_2$H$^+$ and NH$_3$ (Fig.~\ref{FigKIDARoueffAbuFrac}). Instead, both N$_2$ and ${\rm N^{15}NH^+}$ actually join ${\rm ^{15}NNH^+}$ to become less enriched in $^{15}$N( as compared to the nominal ratio of 220) late in cloud evolution.

From the left panel of Figure~\ref{FigKIDARoueffAbuFrac} it can be noted that the overall ammonia abundance is somewhat depressed, due of the lower efficiency of reactions (RN17), the step initiating all ammonia formation. Since electron transfer from amines (NH$_2$, NH$_3$) was a significant destruction channel for H$^+$ in the W12 model, this lower efficiency results in a higher H$^+$ abundance from $2\times10^4$ -- $10^6$~yrs, causing a faster conversion of $o$-H$_2$ to $p$-H$_2$, observed as an earlier drop in the $o$-H$_2$ abundance.

In the right panel of Figure~\ref{FigKIDARoueffAbuFrac}, ammonia does not only show less $^{15}$N enhancement at early and late times than predicted by previous models \citep{RodgersCharnley08,Wirstrom12}, but shows a significantly more pronounced depletion (reaching NH$_3$/$^{15}$NH$_3>2800$) and no enhancement at all (Fig~\ref{FigKIDARoueffAbuFrac}). This is both due to the faster conversion of $o$-H$_2$ to $p$-H$_2$ -- which shifts the onset of $^{15}$N depletion in NH$_3$ to earlier times -- as well as the prolonged presence of molecular nitrogen, delaying the late enhancement as discussed above.

\subsection{Main reservoir of N and $^{15}$N}
It was discussed by \citep{RodgersCharnley08} that a substantial fraction of N in atomic form is required to produce substantial $^{15}$N-fractionation. However, for the revised reaction network where reaction (RF6) is prohibited from converting $^{15}$N from atomic to molecular form, and ammonia is predicted to be predominantly depleted, the major effect of a low initial fraction of atomic nitrogen is on the nitriles (see left hand side of the network in Fig~\ref{FigFracNetw}). The overall abundance of HCN and HNC becomes lower, CN is only significantly enhanced in $^{15}$N just before it reaches its abundance peak and freeze out, but then the enhancement is also transferred to HCN and HNC. $^{15}$N depletion in ammonia is not affected by a low atomic fraction.

On the other hand, starting out with a completely atomic nitrogen reservoir will suppress ammonia and N$_2$H$^+$ abundances, and only after $\sim10^4$~yrs a significant molecular nitrogen abundance is built up via the reaction N~+~NO. 
This has the effect of allowing some $^{15}$N enhancement in ammonia (see right hand side of Fig.~\ref{FigFracNetw}) before depletion sets in when the $o$-H$_2$ abundance drops after $\sim10^5$~yrs (see Fig.~\ref{FigKIDARoueffAbuFrac}) -- and the depletion will not be as pronounced as if starting out with molecular nitrogen. Fractionation in N$_2$H$^+$ will follow that of N$_2$ even more closely because of less frequent interaction between them, and never be significantly depleted or enhanced. 
The high atomic nitrogen abundance will only marginally increase the CN abundance, and therefore nitrile fractionation remains largely the same as in the nominal model.

We also note that observed isotopic ratios of course will depend on which form the initial reservoir of $^{15}$N takes - if it deviates from the atomic to molecular ratio of $^{14}$N. With 50 per cent of $^{14}$N in atomic form as in our standard model, a dominating atomic $^{15}$N reservoir results in $^{15}$N enhanced CN, HCN and HNC at all times, while NH$_3$ and N$_2$H$^+$ will show $^{15}$N depletion at different levels scaling with the initial $^{15}$N/N$^{15}$N ratio. In less dense regions where the ambient UV field can penetrate, the atomic reservoir is predicted to be enhanced in $^{15}$N by isotope-selective photodissociation of N$_2$  \citep{Heays14}. However, if such a precursor cloud retained this enhancement through collapse, the enhancement could also be common in the cold, dark environments of a pre-stellar core, although the process itself is inactive here. Detailed modelling of this evolutionary process will be necessary to ascertain its effectiveness in producing high atomic $^{15}$N fractions in cold, dark environments.

\section{Conclusions}
 
In summary, when the revised $^{15}$N fractionation rates advocated by \citet{Roueff15} are incorporated into the model of \citet{Wirstrom12} the effect of suppressing $^{15}$N enrichments in the nitriles can be offset by including fractionation in the neutral exchange process (RF9) involving atomic nitrogen and the CN radical. On the other hand, ammonia is predicted to never become enriched in $^{15}$N but can still show depletion, whereas the ${\rm N_2H^+}$ isotopologues closely follow the isotope ratio of molecular nitrogen: can retain enrichment but still show no depletion.  When the nitrogen chemistry is updated with reaction rates from the KIDA database, the effects on the $^{15}$N fractionation of ${\rm N_2H^+}$ and ${\rm NH_3}$ are largely unchanged. However, we find that no enrichment occurs in HCN or HNC, contrary to observational evidence in both dark clouds and comets, although it can still be significant in CN itself. With CN at such low abundances, the updated model also falls short of explaining the bulk $^{15}$N enhancements observed in primitive materials. Even if the initial isotopic abundance ratio $^{14}$N/$^{15}$N of the solar nebula was lower, as suggested by some observations \citep{AdandeZiurys12,HilyBlantetal17}, CN would need to be solely responsible for the localised extreme $^{15}$N enhancements in meteoritic insoluble organic matter \citep[$^{14}$N/$^{15}$N$\geq$65, cf.][]{Busemann06}.
 
The reaction barriers of (RF6) and (RF7) are evaluated based on a linear geometry in R15, and it has been pointed out that other approach angles on the potential energy surfaces still might allow the reaction to proceed (G. Nyman, priv. comm.), which could reintroduce the $^{15}$N enhancement in nitriles. However, assuming that this possibility can be neglected, alternative fractionating  bimolecular reactions are necessary to reproduce observations, perhaps involving neutral processes.
It is unclear at this time which may be most viable and so laboratory studies are urgently needed.
Insights into possible alternative fractionation pathways may come from future observational studies that target  all the relevant molecules -- HCN, HNC, CN, ${\rm N_2H^+}$ and ${\rm NH_3}$ -- in individual sources; thus far, these data only exist for  L1544 and Barnard 1 \citep[e.g.][]{Wirstrom16_FM}.  Mapping of these molecules may also provide insight into fractionation mechanisms. 
One interesting possibility is that ${\rm ^{15}N_2H^+}$ could be a significant reservoir of $^{15}$N; recent spectroscopic data mean that searches for this molecule are now possible \citep{Dore17}.
 

\section*{Acknowledgements}
E.S.W. acknowledges generous support from the Swedish National Space Board.
S.B.C. was supported by NASA's Origins of Solar Systems and Emerging Worlds Programs.



\bibliographystyle{mnras}
\bibliography{./references} 

\newcommand{\noopsort}[1]{}
\begin{thebibliography}{}
\makeatletter
\relax
\def\mn@urlcharsother{\let\do\@makeother \do\$\do\&\do\#\do\^\do\_\do\%\do\~}
\def\mn@doi{\begingroup\mn@urlcharsother \@ifnextchar [ {\mn@doi@}
  {\mn@doi@[]}}
\def\mn@doi@[#1]#2{\def\@tempa{#1}\ifx\@tempa\@empty \href
  {http://dx.doi.org/#2} {doi:#2}\else \href {http://dx.doi.org/#2} {#1}\fi
  \endgroup}
\def\mn@eprint#1#2{\mn@eprint@#1:#2::\@nil}
\def\mn@eprint@arXiv#1{\href {http://arxiv.org/abs/#1} {{\tt arXiv:#1}}}
\def\mn@eprint@dblp#1{\href {http://dblp.uni-trier.de/rec/bibtex/#1.xml}
  {dblp:#1}}
\def\mn@eprint@#1:#2:#3:#4\@nil{\def\@tempa {#1}\def\@tempb {#2}\def\@tempc
  {#3}\ifx \@tempc \@empty \let \@tempc \@tempb \let \@tempb \@tempa \fi \ifx
  \@tempb \@empty \def\@tempb {arXiv}\fi \@ifundefined
  {mn@eprint@\@tempb}{\@tempb:\@tempc}{\expandafter \expandafter \csname
  mn@eprint@\@tempb\endcsname \expandafter{\@tempc}}}

\bibitem[\protect\citeauthoryear{{Adams} \& {Smith}}{{Adams} \&
  {Smith}}{1981}]{AdamsSmith81}
{Adams} N.~G.,  {Smith} D.,  1981, \mn@doi [\apjl] {10.1086/183604}, \href
  {http://adsabs.harvard.edu/abs/1981ApJ...247L.123A} {247, L123}

\bibitem[\protect\citeauthoryear{{Adande} \& {Ziurys}}{{Adande} \&
  {Ziurys}}{2012}]{AdandeZiurys12}
{Adande} G.~R.,  {Ziurys} L.~M.,  2012, \mn@doi [\apj]
  {10.1088/0004-637X/744/2/194}, \href
  {http://adsabs.harvard.edu/abs/2012ApJ...744..194A} {744, 194}

\bibitem[\protect\citeauthoryear{{Adande} \& {et al.}}{{Adande} \& {et
  al.}}{2017}]{Adande17}
{Adande} G.~R.,  {et al.} 2017, in prep.

\bibitem[\protect\citeauthoryear{{Bizzocchi}, {Caselli}, {Leonardo}  \&
  {Dore}}{{Bizzocchi} et~al.}{2013}]{Bizzocchi13}
{Bizzocchi} L.,  {Caselli} P.,  {Leonardo} E.,   {Dore} L.,  2013, \mn@doi
  [\aap] {10.1051/0004-6361/201321276}, \href
  {http://adsabs.harvard.edu/abs/2013A%26A...555A.109B} {555, A109}

\bibitem[\protect\citeauthoryear{{Bockel{\'e}e-Morvan}
  et~al.,}{{Bockel{\'e}e-Morvan} et~al.}{2015}]{BockeleeMorvan15}
{Bockel{\'e}e-Morvan} D.,  et~al., 2015, \ssr

\bibitem[\protect\citeauthoryear{{Busemann}, {Young}, {O'D.~Alexander},
  {Hoppe}, {Mukhopadhyay}  \& {Nittler}}{{Busemann} et~al.}{2006}]{Busemann06}
{Busemann} H.,  {Young} A.~F.,  {O'D.~Alexander} C.~M.,  {Hoppe} P.,
  {Mukhopadhyay} S.,   {Nittler} L.~R.,  2006, \mn@doi [Science]
  {10.1126/science.1123878}, \href
  {http://adsabs.harvard.edu/abs/2006Sci...312..727B} {312, 727}

\bibitem[\protect\citeauthoryear{{Charnley} \& {Rodgers}}{{Charnley} \&
  {Rodgers}}{2002}]{CharnleyRodgers02}
{Charnley} S.~B.,  {Rodgers} S.~D.,  2002, \mn@doi [\apjl] {10.1086/340484},
  \href {http://adsabs.harvard.edu/abs/2002ApJ...569L.133C} {569, L133}

\bibitem[\protect\citeauthoryear{{Daranlot}, {Hincelin}, {Bergeat}, {Costes},
  {Loison}, {Wakelam}  \& {Hickson}}{{Daranlot} et~al.}{2012}]{Daranlot12}
{Daranlot} J.,  {Hincelin} U.,  {Bergeat} A.,  {Costes} M.,  {Loison} J.-C.,
  {Wakelam} V.,   {Hickson} K.~M.,  2012, \mn@doi [Proceedings of the National
  Academy of Science] {10.1073/pnas.1200017109}, \href
  {http://adsabs.harvard.edu/abs/2012PNAS..10910233D} {109, 10233}

\bibitem[\protect\citeauthoryear{{Dislaire}, {Hily-Blant}, {Faure}, {Maret},
  {Bacmann}  \& {Pineau Des For{\^e}ts}}{{Dislaire} et~al.}{2012}]{Dislaire12}
{Dislaire} V.,  {Hily-Blant} P.,  {Faure} A.,  {Maret} S.,  {Bacmann} A.,
  {Pineau Des For{\^e}ts} G.,  2012, \mn@doi [\aap]
  {10.1051/0004-6361/201117765}, \href
  {http://adsabs.harvard.edu/abs/2012A%26A...537A..20D} {537, A20}

\bibitem[\protect\citeauthoryear{{Dore}, {Bizzocchi}, {Wirstr{\"o}m}, {Degli
  Esposti}, {Tamassia}  \& {Charnley}}{{Dore} et~al.}{2017}]{Dore17}
{Dore} L.,  {Bizzocchi} L.,  {Wirstr{\"o}m} E.~S.,  {Degli Esposti} C.,
  {Tamassia} F.,   {Charnley} S.~B.,  2017, \mn@doi [\aap]
  {10.1051/0004-6361/201629725}, \href
  {http://adsabs.harvard.edu/abs/2017A%26A...604A..26D} {604, A26}

\bibitem[\protect\citeauthoryear{{Fontani}, {Caselli}, {Palau}, {Bizzocchi}  \&
  {Ceccarelli}}{{Fontani} et~al.}{2015}]{Fontani15}
{Fontani} F.,  {Caselli} P.,  {Palau} A.,  {Bizzocchi} L.,   {Ceccarelli} C.,
  2015, \mn@doi [\apjl] {10.1088/2041-8205/808/2/L46}, \href
  {http://adsabs.harvard.edu/abs/2015ApJ...808L..46F} {808, L46}

\bibitem[\protect\citeauthoryear{{Fukutani} \& {Sugimoto}}{{Fukutani} \&
  {Sugimoto}}{2013}]{FukutaniSugimoto13}
{Fukutani} K.,  {Sugimoto} T.,  2013, \mn@doi [Progress In Surface Science]
  {10.1016/j.progsurf.2013.09.001}, \href
  {http://adsabs.harvard.edu/abs/2013PrSS...88..279F} {88, 279}

\bibitem[\protect\citeauthoryear{{F{\"u}ri} \& {Marty}}{{F{\"u}ri} \&
  {Marty}}{2015}]{FuriMarty15}
{F{\"u}ri} E.,  {Marty} B.,  2015, \mn@doi [Nature Geoscience]
  {10.1038/ngeo2451}, \href {http://adsabs.harvard.edu/abs/2015NatGe...8..515F}
  {8, 515}

\bibitem[\protect\citeauthoryear{{Gerin}, {Marcelino}, {Biver}, {Roueff},
  {Coudert}, {Elkeurti}, {Lis}  \& {Bockel{\'e}e-Morvan}}{{Gerin}
  et~al.}{2009}]{Gerin09}
{Gerin} M.,  {Marcelino} N.,  {Biver} N.,  {Roueff} E.,  {Coudert} L.~H.,
  {Elkeurti} M.,  {Lis} D.~C.,   {Bockel{\'e}e-Morvan} D.,  2009, \mn@doi
  [\aap] {10.1051/0004-6361/200911759}, \href
  {http://adsabs.harvard.edu/abs/2009A%26A...498L...9G} {498, L9}

\bibitem[\protect\citeauthoryear{{Grozdanov}, {McCarroll}  \&
  {Roueff}}{{Grozdanov} et~al.}{2016}]{Grozdanov16}
{Grozdanov} T.~P.,  {McCarroll} R.,   {Roueff} E.,  2016, \mn@doi [\aap]
  {10.1051/0004-6361/201628092}, \href
  {http://adsabs.harvard.edu/abs/2016A%26A...589A.105G} {589, A105}

\bibitem[\protect\citeauthoryear{{Heays}, {Visser}, {Gredel}, {Ubachs},
  {Lewis}, {Gibson}  \& {van Dishoeck}}{{Heays} et~al.}{2014}]{Heays14}
{Heays} A.~N.,  {Visser} R.,  {Gredel} R.,  {Ubachs} W.,  {Lewis} B.~R.,
  {Gibson} S.~T.,   {van Dishoeck} E.~F.,  2014, \mn@doi [\aap]
  {10.1051/0004-6361/201322832}, \href
  {http://adsabs.harvard.edu/abs/2014A%26A...562A..61H} {562, A61}

\bibitem[\protect\citeauthoryear{{Hily-Blant}, {Bonal}, {Faure}  \&
  {Quirico}}{{Hily-Blant} et~al.}{2013a}]{Hily-Blant13a}
{Hily-Blant} P.,  {Bonal} L.,  {Faure} A.,   {Quirico} E.,  2013a, \mn@doi
  [\icarus] {10.1016/j.icarus.2012.12.015}, \href
  {http://adsabs.harvard.edu/abs/2013Icar..223..582H} {223, 582}

\bibitem[\protect\citeauthoryear{{Hily-Blant}, {Pineau des For{\^e}ts},
  {Faure}, {Le Gal}  \& {Padovani}}{{Hily-Blant} et~al.}{2013b}]{Hily-Blant13b}
{Hily-Blant} P.,  {Pineau des For{\^e}ts} G.,  {Faure} A.,  {Le Gal} R.,
  {Padovani} M.,  2013b, \mn@doi [\aap] {10.1051/0004-6361/201321364}, \href
  {http://adsabs.harvard.edu/abs/2013A%26A...557A..65H} {557, A65}

\bibitem[\protect\citeauthoryear{{Hily-Blant}, {Magalhaes}, {Kastner}, {Faure},
  {Forveille}  \& {Qi}}{{Hily-Blant} et~al.}{2017}]{HilyBlantetal17}
{Hily-Blant} P.,  {Magalhaes} V.,  {Kastner} J.,  {Faure} A.,  {Forveille} T.,
   {Qi} C.,  2017, \mn@doi [\aap] {10.1051/0004-6361/201730524}, \href
  {http://adsabs.harvard.edu/abs/2017A%26A...603L...6H} {603, L6}

\bibitem[\protect\citeauthoryear{{Le Teuff}, {Millar}  \& {Markwick}}{{Le
  Teuff} et~al.}{2000}]{LeTeuff00}
{Le Teuff} Y.~H.,  {Millar} T.~J.,   {Markwick} A.~J.,  2000, \mn@doi [\aaps]
  {10.1051/aas:2000265}, \href
  {http://adsabs.harvard.edu/abs/2000A%26AS..146..157L} {146, 157}

\bibitem[\protect\citeauthoryear{{Lis}, {Wootten}, {Gerin}  \& {Roueff}}{{Lis}
  et~al.}{2010}]{Lis10}
{Lis} D.~C.,  {Wootten} A.,  {Gerin} M.,   {Roueff} E.,  2010, \mn@doi [\apjl]
  {10.1088/2041-8205/710/1/L49}, \href
  {http://adsabs.harvard.edu/abs/2010ApJ...710L..49L} {710, L49}

\bibitem[\protect\citeauthoryear{{Novotn{\'y}} et~al.,}{{Novotn{\'y}}
  et~al.}{2014}]{Novotny14}
{Novotn{\'y}} O.,  et~al., 2014, \mn@doi [\apj] {10.1088/0004-637X/792/2/132},
  \href {http://adsabs.harvard.edu/abs/2014ApJ...792..132N} {792, 132}

\bibitem[\protect\citeauthoryear{{Rodgers} \& {Charnley}}{{Rodgers} \&
  {Charnley}}{2008a}]{RodgersCharnley08}
{Rodgers} S.~D.,  {Charnley} S.~B.,  2008a, \mn@doi [\mnras]
  {10.1111/j.1745-3933.2008.00431.x}, \href
  {http://adsabs.harvard.edu/abs/2008MNRAS.385L..48R} {385, L48}

\bibitem[\protect\citeauthoryear{{Rodgers} \& {Charnley}}{{Rodgers} \&
  {Charnley}}{2008b}]{RodgersCharnley08_ApJ}
{Rodgers} S.~D.,  {Charnley} S.~B.,  2008b, \mn@doi [\apj] {10.1086/592195},
  \href {http://adsabs.harvard.edu/abs/2008ApJ...689.1448R} {689, 1448}

\bibitem[\protect\citeauthoryear{{Roueff}, {Loison}  \& {Hickson}}{{Roueff}
  et~al.}{2015}]{Roueff15}
{Roueff} E.,  {Loison} J.~C.,   {Hickson} K.~M.,  2015, \mn@doi [\aap]
  {10.1051/0004-6361/201425113}, \href
  {http://adsabs.harvard.edu/abs/2015A%26A...576A..99R} {576, A99}

\bibitem[\protect\citeauthoryear{{Savage} \& {Sembach}}{{Savage} \&
  {Sembach}}{1996}]{SavageSembach96}
{Savage} B.~D.,  {Sembach} K.~R.,  1996, \mn@doi [\araa]
  {10.1146/annurev.astro.34.1.279}, \href
  {http://adsabs.harvard.edu/abs/1996ARA%26A..34..279S} {34, 279}

\bibitem[\protect\citeauthoryear{{Shinnaka}, {Kawakita}, {Jehin}, {Decock},
  {Hutsem{\'e}kers}, {Manfroid}  \& {Arai}}{{Shinnaka}
  et~al.}{2016}]{Shinnaka16}
{Shinnaka} Y.,  {Kawakita} H.,  {Jehin} E.,  {Decock} A.,  {Hutsem{\'e}kers}
  D.,  {Manfroid} J.,   {Arai} A.,  2016, \mn@doi [\mnras]
  {10.1093/mnras/stw2410}, \href
  {http://adsabs.harvard.edu/abs/2016MNRAS.462S.195S} {462, S195}

\bibitem[\protect\citeauthoryear{{Terzieva} \& {Herbst}}{{Terzieva} \&
  {Herbst}}{2000}]{TerzievaHerbst00}
{Terzieva} R.,  {Herbst} E.,  2000, \mn@doi [\mnras]
  {10.1046/j.1365-8711.2000.03618.x}, \href
  {http://adsabs.harvard.edu/abs/2000MNRAS.317..563T} {317, 563}

\bibitem[\protect\citeauthoryear{{Wakelam} et~al.,}{{Wakelam}
  et~al.}{2010}]{Wakelam10b}
{Wakelam} V.,  et~al., 2010, \mn@doi [\ssr] {10.1007/s11214-010-9712-5}, \href
  {http://adsabs.harvard.edu/abs/2010SSRv..156...13W} {156, 13}

\bibitem[\protect\citeauthoryear{{Wakelam}, {Smith}, {Loison}, {Talbi},
  {Klippenstein}, {Bergeat}, {Geppert}  \& {Hickson}}{{Wakelam}
  et~al.}{2013}]{Wakelam13}
{Wakelam} V.,  {Smith} I.~W.~M.,  {Loison} J.-C.,  {Talbi} D.,  {Klippenstein}
  S.~J.,  {Bergeat} A.,  {Geppert} W.~D.,   {Hickson} K.~M.,  2013, preprint,
  \href {http://adsabs.harvard.edu/abs/2013arXiv1310.4350W} {} (\mn@eprint
  {arXiv} {1310.4350})

\bibitem[\protect\citeauthoryear{{Wakelam} et~al.,}{{Wakelam}
  et~al.}{2015}]{Wakelam15}
{Wakelam} V.,  et~al., 2015, \mn@doi [\apjs] {10.1088/0067-0049/217/2/20},
  \href {http://adsabs.harvard.edu/abs/2015ApJS..217...20W} {217, 20}

\bibitem[\protect\citeauthoryear{{Wirstr{\"o}m}, {Charnley}, {Cordiner}  \&
  {Milam}}{{Wirstr{\"o}m} et~al.}{2012}]{Wirstrom12}
{Wirstr{\"o}m} E.~S.,  {Charnley} S.~B.,  {Cordiner} M.~A.,   {Milam} S.~N.,
  2012, \mn@doi [\apjl] {10.1088/2041-8205/757/1/L11}, \href
  {http://adsabs.harvard.edu/abs/2012ApJ...757L..11W} {757, L11}

\bibitem[\protect\citeauthoryear{{Wirstr{\"o}m}, {Adande}, {Milam}, {Charnley}
  \& {Cordiner}}{{Wirstr{\"o}m} et~al.}{2016}]{Wirstrom16_FM}
{Wirstr{\"o}m} E.~S.,  {Adande} G.,  {Milam} S.~N.,  {Charnley} S.~B.,
  {Cordiner} M.~A.,  2016, \mn@doi [IAU Focus Meeting]
  {10.1017/S1743921316003033}, \href
  {http://adsabs.harvard.edu/abs/2016IAUFM..29A.271W} {29, 271}

\bibitem[\protect\citeauthoryear{{van Kooten} et~al.,}{{van Kooten}
  et~al.}{2017}]{vanKooten17}
{van Kooten} E.~M.~M.~E.,  et~al., 2017, \mn@doi [\gca]
  {10.1016/j.gca.2017.02.002}, \href
  {http://adsabs.harvard.edu/abs/2017GeCoA.205..119V} {205, 119}

\makeatother
\end{thebibliography}








\bsp	
\label{lastpage}
\end{document}